\documentclass{KapProc} % Computer Modern font calls

\usepackage{psfig}

\kluwerbib
\begin{document}

\articletitle{High-mass X-ray binaries and OB-runaway stars}

%\articlesubtitle{This is an Article Subtitle}

\author{Lex Kaper}
\affil{Astronomical Institute, University of Amsterdam,\\
Kruislaan 403, 1098 SJ Amsterdam, The Netherlands}
\email{lexk@astro.uva.nl}

\begin{abstract}
High-mass X-ray binaries (HMXBs) represent an important phase in the
evolution of massive binary systems.  HMXBs provide unique diagnostics
to test massive-star evolution, to probe the physics of
radiation-driven winds, to study the process of mass accretion, and to
measure fundamental parameters of compact objects.  As a consequence
of the supernova explosion that produced the neutron star (or black
hole) in these systems, HMXBs have high space velocities and thus are
runaways. Alternatively, OB-runaway stars can be ejected from a
cluster through dynamical interactions. Observations obtained with the
{\it Hipparcos} satellite indicate that both scenarios are at
work. Only for a minority of the OB runaways (and HMXBs) a wind bow
shock has been detected.  This might be explained by the varying local
conditions of the interstellar medium.
\end{abstract}


\section{The evolution of massive binaries}

In a high-mass X-ray binary (HMXB) a massive, OB-type star is orbited
by a compact, accreting X-ray source: a neutron star or a black
hole. Mass is accreted either from the stellar wind of the massive
companion (resulting in a modest X-ray luminosity), or flows with a
higher rate to the compact star when the massive star fills its
Roche lobe. In the latter case a much higher X-ray luminosity is
achieved. Two classes of HMXBs can be distinguished based on the
nature of the massive star: (i) the OB-supergiant systems, and (ii)
the Be/X-ray binaries (for a recent catalogue of HMXBs, see Liu et
al. 2000). The first class consists of the most massive systems, some
of them containing a black hole (e.g. Cyg~X-1). The second class
comprises the majority ($\sim$80~\%) of the HMXBs; most of them are
detected as X-ray transients. The transient character is explained by
the periodic increase in X-ray luminosity when the neutron star, in
its eccentric orbit, passes through periastron and accretes from the
dense equatorial Be-star disk. In the remainder of this paper, we will
concentrate on the OB-supergiant systems.

HMXBs are the descendants of massive binaries in which the initially
most massive star (the primary) has become a neutron star or a black
hole (Van den Heuvel \& Heise 1972; for extensive reviews on binary
evolution, see e.g.\ Van den Heuvel 1993, Vanbeveren et
al. 1998). That the system remains bound after the supernova is due to
a phase of mass transfer that occurs in the system when the primary
grows larger than its Roche lobe. As a consequence, the secondary
becomes the most massive of the two, so that with the supernova
explosion of the primary less than half of the total system mass is
lost and the system remains bound (Boersma 1961). It is assumed here
that the supernova explosion is symmetric; in case of an asymmetry,
the additional kick exerted on the compact remnant can cause the
disruption of the system. Anyway, due to the loss of material (and
momentum), the system gets a substantial ``runaway'' velocity (Blaauw
1961): on the order of 50~km~s$^{-1}$ for the most massive systems
(and about 15~km~s$^{-1}$ for the less massive Be/X-ray binaries, cf.\
Van den Heuvel et al. 2001).

As long as the secondary is a massive main sequence star,
accretion of its relatively tenuous stellar wind onto the compact
companion does not result in an observable X-ray flux. Only when the
secondary becomes a supergiant and starts filling its Roche
lobe -- this scenario refers to Case~B binary evolution; in Case
A the secondary fills its Roche lobe already on the main sequence --
the accretion flow is dense enough to power a strong X-ray source; now
the system has become a HMXB. This phase lasts for a relatively short
period of time ($\sim$10,000 year): as soon as the Roche-lobe overflow
commences, the orbit will shrink leading to an even higher mass
transfer rate and a further tightening of the orbit. At some point, the
X-ray source will be completely swamped with material optically thick
in X-rays and/or penetrate the mantle of the secondary, followed by a
rapid spiral-in. When also the secondary explodes as a supernova, a
neutron-star binary (like the Hulse-Taylor binary pulsar) or two
single neutron stars remain. The detailed evolution of massive
binaries is still a matter of debate (e.g. Wellstein \& Langer 1999,
and several contributions to these proceedings).

\begin{table}[ht]
\caption[High-mass X-ray binaries with OB supergiant
companion]{High-mass X-ray binaries with OB supergiant companion in
the Milky Way and the Magellanic Clouds (ordered according to right
ascension). The name corresponds to the X-ray source, the spectral
type to the OB supergiant. For the systems hosting an X-ray pulsar the
masses of both binary components can be measured (given an estimate of
the inclination of the system). The last two systems contain a
black-hole candidate. The radius of the OB supergiant can be
determined in case X-ray eclipses are observed. The system parameters
were taken from Van Kerkwijk et al. (1995), except for 2S0114+650
(Reig et al. 1996), Vela~X-1 (Barziv et al. 2001), GX301-2 (Kaper \&
Najarro 2001), 4U1700-37 (Heap \& Corcoran 1992), 4U1907+09 (Van
Kerkwijk et al. 1989), LMC~X-1 (Hutchings et al. 1987), Cyg~X-1
(Herrero et al. 1995). The rapid X-ray pulsars are found in Roche-lobe
overflow systems. The space velocity (or only the radial component)
has been measured for some HMXBs and has been corrected for the solar
peculiar motion and differential galactic rotation.}
\begin{tabular*}{\textwidth}{@{\extracolsep{\fill}}llcrccc}
\sphline
Name & Sp.\ Type & M$_{\rm OB}$ & R & M$_{\rm X}$ & P$_{\rm pulse}$ &
$v_{\rm space}$ \cr
     &           & (M$_{\odot}$) & (R$_{\odot}$) & (M$_{\odot}$) & (s) & 
(km~s$^{-1}$) \cr 
\sphline
2S0114+650 & B1 Ia     & $\sim$16 & & $\sim$1.7 & 860$^a$ & 32 \cr
SMC X-1   & B0 Ib         & 17.2 & 15 & 1.6    & 0.71 & \cr
LMC X-4   & O7 III-IV     & 15.8 & 8  & 1.5    & 13.5 & \cr
Vela~X-1  & B0.5~Ib       & 23.9 & 30 & 1.9    & 283  & 35 \cr
Cen~X-3   & O6.5 II-III   & 18.4 & 11 & 1.1    & 4.84 & 42$^b$ \cr
GX301-2   & B1.5 Ia$^{+}$ & $>$40 &   & $>$1.3 & 696  & $\sim$5$^b$ \cr
4U1538-52 & B0 Iab        & 16.4 & 15 & 1.1    & 529  & 85$^b$ \cr
4U1700-37 & O6.5 Iaf+     & $\sim$52$^c$ & 22 & $\sim$1.8$^c$ &  & 76 \cr
4U1907+09 & early B I     & $>$9 &    & $>$0.7 & 438  & \cr \sphline
%LMC X-3   & B3 V          &      &    & 3.5-10 &      & \cr
LMC X-1   & O7-9 III      &      &    & 4-10   &      & \cr
Cyg X-1   & O9.7 Iab      & 17.8 &    & 10     &      & $\sim$20$^d$ \cr
\sphline
\end{tabular*}
\begin{tablenotes}
$^a$ A spin period of 2.7h is proposed by Corbet et al. (1999); $^b$
Only the radial-velocity component has been measured; $^c$ Rubin et
al. (1996) give 30~M$_{\odot}$ and 2.6~M$_{\odot}$ for the mass of the
O supergiant and X-ray source, respectively; $^d$ The space velocity
of Cyg~X-1 is with respect to the OB-star population in its local
environment.
\end{tablenotes}
\end{table}
%\inxx{captions,table}

In the following we will discuss the observational consequences of the
evolutionary scenario outlined above for the properties of high-mass
X-ray binaries. Firstly, we will compare the stellar parameters of the
OB supergiants in HMXBs to those of single massive stars. These stars
gained a significant fraction of their present mass from the
progenitor of the X-ray source; this matter likely is chemically
enriched. Not only matter, but also angular momentum is
transfered. Therefore, both the internal and the atmospheric
properties of these OB stars might differ from those of single massive
stars. Secondly, the binarity of these systems provides the
opportunity to measure fundamental parameters (such as the mass) of
neutron stars and black holes.

Thirdly, the impact of the X-ray source on the structure of the
OB-supergiant's wind is discussed. The X-ray source creates a
Str\"{o}mgren zone of high ionization in which the radiative
acceleration of the wind is quenched. This leads to the formation of a
photo-ionization wake trailing the X-ray source in its orbit. In the
most extreme case, only in the X-ray shadow a radiatively driven wind
emerges: a so-called shadow wind.

The last section deals with the runaway nature of HMXBs. Although on
theoretical grounds high space velocities are expected, the
observational proof that HMXBs are runaway objects has just recently
been obtained. {\it Hipparcos} results indicate that both
the binary supernova scenario and the cluster ejection mechanism
produce OB-runaway stars.

\begin{figure}[t]
%Fig. 1
\centerline{\psfig{figure=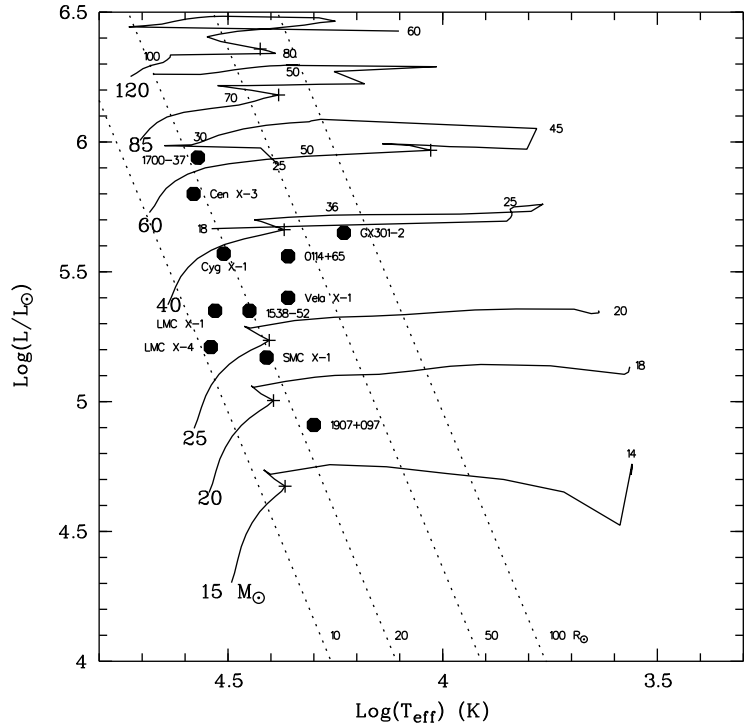,width=10cm,angle=-90}}
\caption[]{The location of the OB supergiants in HMXBs (filled
circles) in the Hertzsprung-Russell diagram. The evolutionary tracks
of Lejeune \& Schaerer (2000) for stars of given mass are drawn in (up
to the phase of core-helium burning). The numbers along the tracks
(smaller font) show the decrease in mass of the star with time due to
mass loss. The plus signs indicate the end of core-hydrogen
burning. The diagonal dotted lines are lines of constant radius.}
\end{figure}

\section{The OB supergiants in HMXBs}

Table~1 gives an overview of the OB-supergiant systems in the Milky
Way and the Magellanic Clouds. The orbital periods range from 1.4
(LMC~X-4) to 41.5~days (GX301-2). A relatively long pulse period
(minutes) indicates that the X-ray source is a neutron star in a
wind-fed system. The Roche-lobe overflow systems (e.g. Cen~X-3) host
rapid X-ray pulsars, most likely surrounded by an accretion disk. In
these systems not only more mass, but also more angular momentum is
transfered to the X-ray source, explaining the high (Eddington) X-ray
luminosity and rapid spin period. 

Figure~1 shows the distribution of the sample of OB supergiants in
HMXBs (Table~1) in the Hertzsprung-Russell diagram. The spectral type
of the OB supergiant is used to determine the effective temperature of
the star and to estimate the interstellar reddening. With the latter,
an estimate of the distance, and the bolometric correction (which also
depends on the spectral type), the luminosity is derived. For several
systems the distance is known with reasonable accuracy, especially the
Magellanic Cloud systems and systems for which a parent OB association
has been identified (\S~5). For comparison, the evolutionary tracks of
(single) massive stars are drawn in (Lejeune \& Schaerer 2000), which
take into account the stellar-wind mass loss. The mass on the zero-age
main sequence is indicated (and the reduced mass due to mass loss is
displayed at some locations along the tracks).

\begin{figure}[t]
%Fig. 2
\centerline{\psfig{figure=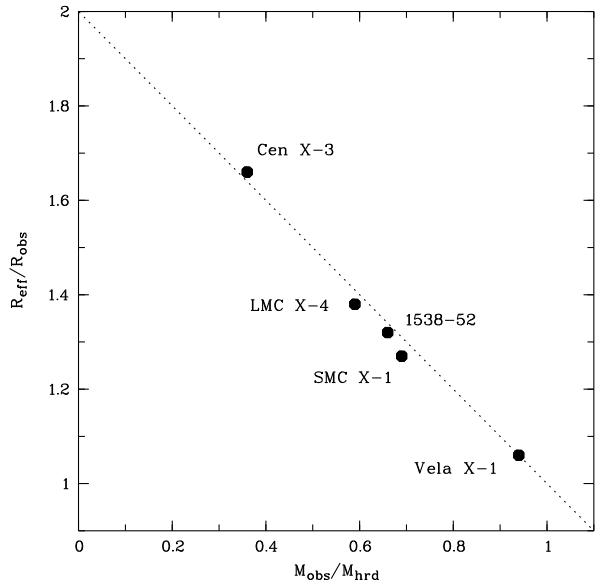,width=8cm,angle=-90}}
\caption[]{The ratio of the observed mass of OB supergiants (in HMXBs
hosting an eclipsing X-ray pulsar) and the mass estimated from
their position in the HRD is compared to the ratio of the effective
temperature radius and the observed radius. The difference between
observed and expected mass and radius seems to be bigger for systems
that are presently in a phase of Roche-lobe overflow. }
\end{figure}

The mass of the OB supergiant in a HMXB can be accurately measured
when the system hosts an X-ray pulsar. With the delay in pulse-arrival
time and the radial-velocity curve of the OB supergiant the orbits of
both stars are determined, and thereby their masses if the orbital
inclination is known. The latter is well constrained if the system is
eclipsing. In those systems, also the radius of the OB supergiant can
be derived with high precision from the duration of the X-ray
eclipse. The masses and radii are listed in Table~1. The mass ratio
sets the size of the Roche lobe (e.g.\ Eggleton 1983); it turns out
that the measured radii of the OB supergiants are in very good
agreement with the estimated size of the Roche lobe.

Earlier studies (e.g.\ Conti 1978, Rappaport \& Joss 1983) suggested
that the OB supergiants in HMXBs are too luminous for their mass. This
is clearly demonstrated in Figure~1: e.g.\ the O6.5 giant companion of
Cen~X-3 has a mass of 19~M$_{\odot}$, while its luminosity corresponds
to that of a star of more than 50~M$_{\odot}$. Besides this, the
radius corresponding to the luminosity and effective temperature
(R$_{\rm eff}$) is larger than its measured (Roche-lobe) radius. Thus,
apart from being undermassive, the OB supergiants in HMXBs also seem
to be {\it undersized} for their luminosity and temperature. Figure~2
shows that the OB supergiants (in systems with an eclipsing X-ray
pulsar) with the largest deviation in radius, are also the most
``undermassive'' ones. The wind-fed system 4U1700-37 is eclipsing, but the
X-ray source is not pulsating; the measured radius of its O6.5~Iaf
companion is in good agreement with R$_{\rm eff}$, but the
supergiant's mass cannot be accurately determined. For a similar reason,
the non-eclipsing HMXBs hosting an X-ray pulsar are not included in
Fig.~2, because the radius is not well known.
 
However, the effective temperature of a star is
determined by its luminosity and size. Apparently, the spectral type
-- effective temperature calibration of single stars cannot be applied
to the OB supergiants in HMXBs (presuming that the distances to these
objects are not systematically too large). For the five objects in
Fig.~2, the deviation in radius and mass is worst for the stars with
earliest spectral type, but this might be a coincidence
(e.g. 4U1700-37 has an O6.5 companion, but has R$_{\rm eff}$/R$_{\rm
obs}$ close to one). The discrepancy between R$_{\rm eff}$ and R$_{\rm
obs}$ can be removed by systematically increasing T$_{\rm eff}$ by
10-25~\%, while keeping the luminosity constant (the bolometric
correction does not change very much). But this does not explain why
the stars are undermassive.

Probably the trend observed in Fig.~2 is related to the phenomenon of
Roche-lobe overflow. The OB star tries to become a supergiant, but at
some point reaches its critical Roche lobe and starts to transfer mass
to its companion. While the luminosity of the star is determined by
the core (which does not notice much of what is happening to the
outer mantle), the star likes to be bigger than allowed by its Roche
lobe and is peeled off. This would explain why systems like Vela~X-1
(and 4U1700-37, GX301-2) which are wind-fed X-ray sources not (yet)
experiencing Roche-lobe overflow, are not showing large discrepancies
in mass and radius. 

Regarding the chemical enrichment of the massive stars in HMXBs, a
thorough analysis is still lacking. The ultraviolet spectrum of
HD77581, the companion of Vela~X-1, shows very strong N~{\sc v} and
Al~{\sc iii} resonance lines, typical for nitrogen enhanced B stars,
the BN stars (Kaper et al. 1993). Atmospheric models of HDE226868, the
O supergiant companion of Cyg~X-1, by Herrero et
al. (1995) indicate a helium abundance higher than expected for its
evolutionary state (a helium discrepancy, assuming single star
evolution). However, according to Herrero et al. (1992) also rapid
rotators and very luminous stars, of which several are single, show
helium enhancement. Blaauw (1993) and Hoogerwerf et al. (2000) used
these findings to suggest that the helium enhancement and rapid
rotation are observational characteristics of OB runaways;
unfortunately, we are still dealing with small-number statistics.

\section{Neutron stars, black holes, and gamma-ray bursts} 

The accurate measurement of neutron-star masses is essential
for our understanding of the equation of state (EOS) of matter at
supra-nuclear densities, and of the mechanism of core collapse of
massive stars. The EOS can, so far, only be studied on the basis of
theoretical models.  These remain very uncertain and subject to hot
dispute. Brown \& Bethe (1994) have strongly argued that
the EOS must be ``soft'', i.e., that matter would be relatively
compressible, due to kaon condensation (kaons are bosons which do not
contribute to the Fermi pressure).  Heavy ion collision
experiments (in Hamburg, DESY) confirm their predictions, but at
densities still quite a bit lower than those appropriate for neutron
stars.  If the theory were correct, one of the astrophysical
implications would be that neutron stars cannot have a mass larger
than 1.55~M$_{\odot}$; for larger masses, the object would collapse
into a black hole.

The masses of X-ray pulsars (Table~1) can also be used to test the
predictions of supernova models. For instance, Timmes et al. (1996)
present model calculations from which they find that Type~II
supernovae (massive, single stars) give a bimodal compact-object mass
distribution, with peaks at 1.28 and 1.73 M$_{\odot}$, while Type~Ib
supernovae (such as produced by stars in binaries, which are stripped
of their envelopes) will produce neutron stars within a small range
around 1.32~M$_{\odot}$. The massive neutron star in Vela~X-1 with a
mass of $1.9 \pm 0.2$~M$_{\odot}$ would be consistent with the second
peak in this distribution (Barziv et al. 2001).

Presently, all accurate mass determinations besides that of Vela~X-1
have been for neutron stars that were almost certainly formed from
Type Ib supernovae and that have accreted little since.  The most
accurate have been derived for the binary radio pulsars.  All of these
are consistent with a small range near $1.35\,M_{\odot}$ (Thorsett \&
Chakrabarty 1999). That this mass is so close to the Chandrasekhar
mass is thought to be due to the maximum mass of the degenerate iron
core, just before supernova. Apparently, the supernova explosion in
this case prevents additional material from accretion onto the
proto-neutron star, thus also preventing the formation of a black
hole. However, the (binary) radio pulsars are likely to originate from
binaries of moderate mass which might not be able to produce massive
neutron stars, nor black holes. This would be consistent with the
observed absence of black holes in Be/X-ray binaries. If massive
neutron stars exist, they are more likely to have formed in systems of
high initial mass. These systems are capable of producing black holes,
thus a fall-back mechanism must exist (because these stars will also
have a degenerate iron core before collapse). The alternative is that
black holes are not produced by normal supernovae but in gamma-ray
bursts (hypernovae), such as SN1998bw/GRB980425 (Galama
et al. 1998, Iwamoto et al. 1998).

Our hypothesis is that when a fall-back mechanism exists, the masses
of neutron stars in the most massive HMXBs will be evenly distributed
over a relatively large mass interval, up to a maximum mass set
by the neutron star equation of state. Alternatively, black holes are
produced in gamma-ray bursts, and massive neutron stars like in
Vela~X-1 are explained by the second of the two peaks in the mass
distribution such as predicted by Timmes et al. (1996).

\section{Interaction between X-ray source and stellar wind}

\begin{figure}[ht]
%Fig. 3
\vspace{8cm}
%Please convert .jpg file into .ps (e.g. with xv) and remove
%percent sign in line below; add percent sign before vspace
%\centerline{\psfig{figure=velax1sketch.ps,width=10cm}}
\caption[]{A sketch of the HMXB Vela~X-1. The X-ray source creates a
Str\"{o}mgren zone in the stellar wind. A photo-ionization wake is
formed at the trailing border of the Str\"{o}mgren zone where the
stagnant flow meets the fast, accelerating wind.}
\end{figure}

The X-ray source ionizes a significant fraction of the ambient stellar
wind (Fig.~3). The presence of a Str\"{o}mgren zone in the system
becomes apparent from the orbital modulation of strong ultraviolet
resonance lines (e.g.\ from N~{\sc v}, Si~{\sc iv}, and C~{\sc iv},
cf.\ Kaper et al. 1993) formed in the stellar wind. When the X-ray
source is in the line of sight ($\phi=0.5$), the blue-shifted
absorption part of the P-Cygni profile is strongly reduced in
strength, because ions like C~{\sc iv} are removed by the X-rays. The
orbital modulation of UV resonance lines is observed in most HMXBs
with OB supergiant companion for which ultraviolet spectra are
available. However, in HD153919/4U1700-37, the system for which
Hatchett \& McCray (1978) originally predicted this effect, the
P-Cygni profiles do not vary with orbital phase. When the
non-monotonic velocity structure of the stellar wind is taken into
account, the absence of the HM-effect in 4U1700-37 can be explained
(Van Loon et al. 2001). For a review on stellar winds in HMXBs, see
Kaper (1998).

A consequence of the high degree of ionization inside the
Str\"{o}mgren zone is that in this region the radiative acceleration
is quenched. The highly ionized plasma becomes effectively transparent
for ultraviolet photons emitted by the OB companion that accelerate
the stellar wind. Due to the revolution of the system, the stagnant
flow inside the ionization zone meets the rapidly accelerating
stellar wind at the trailing border. Here, a strong shock is formed
(Fig.~3), the so-called photo-ionization wake (Blondin et
al. 1990). The relatively large extent of the photo-ionization wake
results in an observable blue-shifted absorption feature in strong
optical lines at late orbital phases (Kaper et al. 1994).

The Roche-lobe overflow systems (SMC~X-1, LMC~X-4, Cen~X-3) contain a
very strong X-ray source. The X-ray luminosity approaches the
bolometric luminosity of the OB companion, so that the X-ray flux
dominates the ionization equilibrium in a large fraction of the
stellar wind. Actually, only in the X-ray shadow behind the OB
companion a normal stellar wind can develop, the shadow wind (Blondin
1994). Ultraviolet spectra of LMC~X-4 (Boroson et al. 1999) clearly
show the presence of a shadow wind in this system. High-resolution UV
spectra of LMC~X-4 obtained with HST/STIS reveal the signature of a
photo-ionization wake, there where the fast
shadow wind rams into the highly ionized Str\"{o}mgren zone (Kaper et
al. 2001).

The spin behaviour of the X-ray pulsar provides information on the
accretion flow in the system. In Roche-lobe overflow systems the X-ray
pulsar is spun up due to the relatively high accretion rate and the
corresponding transfer of angular momentum. The spin-up and spin-down
episodes observed in some wind-fed systems (e.g. Vela~X-1) indicate
that sometimes a small accretion disk is formed, the direction of
rotation depending on details (e.g. instability) of the accretion flow
(Blondin et al. 1990). Variations in the accretion flow are also
apparent from the strong X-ray flaring behaviour observed in HMXBs.

\begin{figure}[ht]
%Fig. 4
\centerline{\psfig{figure=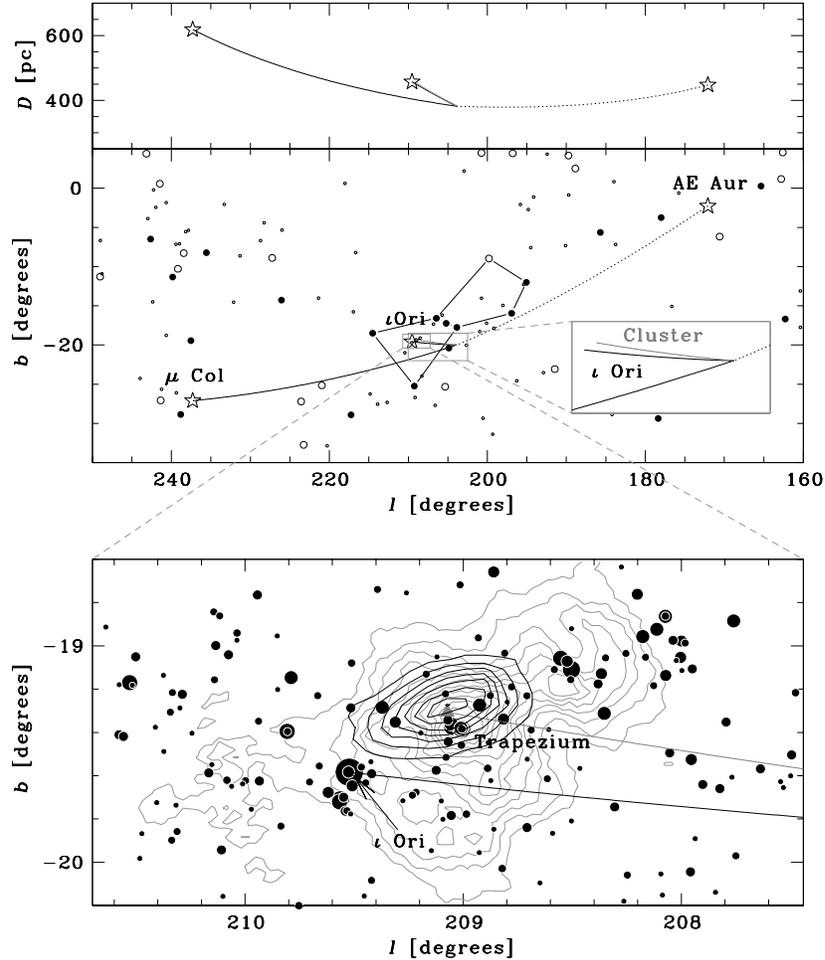,width=11cm}}
\caption[]{Top \& middle: The orbits, calculated back in time, of the
runaways AE~Aur (dotted line) and $\mu$~Col (solid line), and the
binary $\iota$~Ori based on Hipparcos observations. The top panel
shows the distance versus galactic longitude, the middle panel displays
the orbits projected on the sky in galactic coordinates. The stars met
$\sim 2.5$~Myr ago. Filled circles denote O and B stars, open circles
represent other spectral types. Bottom: A blow up of the central
region shows the predicted position of the parent cluster (black
contours) together with all stars in the Tycho Catalogue (ESA 1997) in
the field down to V~=~12.4 mag. The brightest star is $\iota$~Ori;
also the Trapezium is indicated. The grey contours (IRAS 100~$\mu$m
flux) outline the Orion Nebula. Figure taken from Hoogerwerf et
al. (2000).}
\end{figure}

\section{The origin of OB-runaway stars}

About 20\% of the O stars, and a smaller fraction of the B stars, have
a space velocity much higher than observed on average for the OB-star
population in the Milky Way (about 10~km~s$^{-1}$, Stone 1979). Some
OB stars have a space velocity exceeding 100~km~s$^{-1}$. Blaauw
(1961) called these stars runaway stars, because at least for some
of them the reconstructed path through space suggests an origin in a
nearby OB association. How did these massive stars obtain such a high
space velocity? Blaauw proposed that the supernova explosion of a
massive companion in a binary results in a high velocity of the
remaining massive star; the ``modern'' version of this Blaauw scenario
is described in \S~1, which predicts that all HMXBs are runaway
systems. It took several decades to obtain the observational evidence
proving that this scenario actually works (Van Oijen 1989, Van
Rensbergen et al. 1996, Kaper et al. 1997); now, {\it Hipparcos}
measurements have definitely confirmed the runaway nature of HMXBs
(Chevalier \& Ilovaisky 1998, Kaper et al. 1999); Table~1 lists the
derived space velocities.

An alternative scenario for the formation of OB runaways is 
dynamical ejection from a compact cluster (Poveda et al. 1967,
Portegies Zwart 2000). Due to the dynamical interaction between
binaries (and single stars), now and then a massive star is ejected from
an OB association. The probability for ejection is higher in dense
starclusters; given the expansion of OB associations, one expects that
dynamical ejection is most effective in young OB associations. An
impressive example demonstrating the cluster ejection scenario is
provided by the OB runaways AE~Aur and $\mu$~Col. Blaauw \& Morgan
(1954) noticed that these two stars move in almost opposite directions
away from the Ori~OB1 association and have the same kinematical age
(i.e. the travel time from the association to its present location). Using
{\it Hipparcos} observations, Hoogerwerf et al. (2000) managed to
reconstruct the kinematical history of these two runaways in great
detail and could show that both stars and the massive binary system
$\iota$~Ori were at the same place in Ori~OB1 $\sim 2.5$ million years
ago (Fig.~4). The dynamical interaction between two binary systems
apparently led to the disruption of one of them and the subsequent
ejection of the two members (in opposite directions due to momentum
conservation).  

\begin{figure}[t]
%Fig. 5
\centerline{\psfig{figure=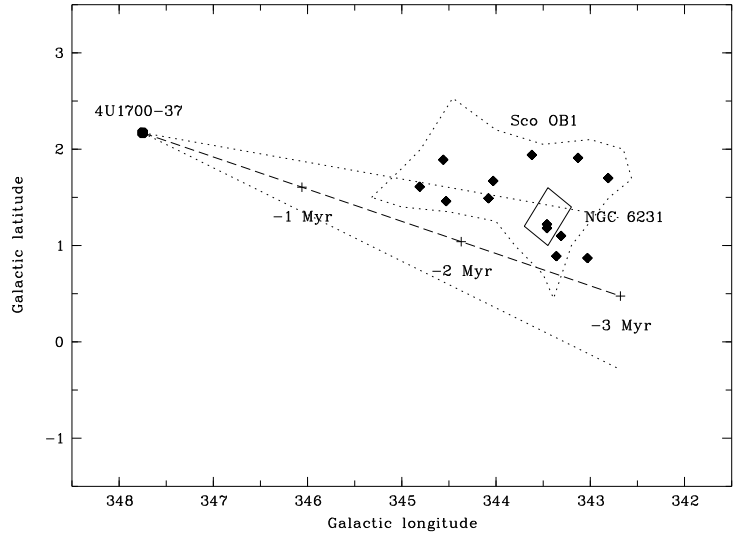,width=11cm,angle=-90}}
\caption[]{The reconstructed path of the runaway HMXB 4U1700-37
intersects with the location of Sco~OB1; the error cone is indicated
by the dotted straight lines. The Hipparcos confirmed members
are shown as filled diamonds. The proper motion of 4U1700-37 is with
respect to the average proper motion of Sco~OB1. The corresponding
kinematical age of 4U1700-37 is $2 \pm 0.5$ million year. The current
angular separation between 4U1700-37 and NGC~6231 (at 2~kpc)
corresponds to a distance of about 150~pc.}
\end{figure}

Based on the {\it Hipparcos} proper motion of HD153919 (4U1700-37),
Ankay et al. (2001) demonstrate that this HMXB very likely originates
from the OB association Sco~OB1 at a distance of 2~kpc (Fig.~5). The
kinematical age of the system is 2$\pm$0.5~Myr, which sets the time of
the supernova producing the compact star in the system. The present
age of Sco~OB1 is 6$\pm$2~Myr (the ``nucleus'' of the association,
NGC~6231, is probably younger). These observations can be used to
derive a constraint on the initial mass of the progenitor of
4U1700-37. The turn-off mass of a cluster of 4~Myr is about
25~M$_{\odot}$. If the system was born as a member of NGC~6231, the
progenitor of 4U1700-37 likely was an even more massive star ($\sim
40$~M$_{\odot}$).

Hoogerwerf et al. (2000) studied the kinematical history of a sample
of nearby runaways (17) and their (candidate) parent OB associations
using {\it Hipparcos} data. A comparison of the kinematical age of the
runaway and the age of the cluster is used to discriminate between the
two formation scenarios. In case of dynamical ejection, the
kinematical age of the runaway should be about equal to the age of the
association. In the binary supernova scenario, the evolution of the
binary system implies that the kinematical age of the runaway is
significantly shorter than the age of the cluster. Furthermore, the OB
runaway will be a blue straggler in the HRD of the cluster, because
the OB runaway is rejuvenated (by gaining mass) during the first phase
of mass transfer in the system. Hoogerwerf et al. conclude that the
two mechanisms produce about equal amounts of runaways. This
conclusion is in agreement with theoretical predictions (e.g. De
Donder et al. 1997, Portegies Zwart 2000).

OB runaways can be used as probes of the interstellar medium; their
stellar wind interacts with the ambient medium reflecting the motion
of the star. Where the ram pressure of wind and interstellar medium
balance, a wind bow shock is formed if the runaway is moving with a
supersonic velocity. Van Buren et al. (1995) detected wind bow shocks
around several OB runaways, with a detection rate of about
30~\%. Kaper et al. (1997) discovered a wind bow shock around the HMXB
Vela~X-1. Huthoff \& Kaper (2001) searched for the presence of wind
bow shocks around other HMXBs, but could not add a significant
detection. The non-detection of wind bow shocks for the majority of
runaways is likely related to the varying physical conditions of the
ISM. For example, inside a hot superbubble the sound speed is about
100~km~s$^{-1}$, so that the runaway would not move supersonically and
a bow shock is not formed. These superbubbles are created by the
combined action of stellar winds and supernovae of the massive stars
in OB associations. Huthoff \& Kaper demonstrate that the OB runaways
with a detected wind bow shock concentrate in regions in between
superbubbles.

\begin{acknowledgments}
It is a pleasure to thank Askin Ankay, Ed van den Heuvel, Gijs
Nelemans, Freek Huthoff, Ronnie Hoogerwerf, Jos de Bruijne, and
Norbert Langer for stimulating discussions. LK is supported by a
fellowship of the Royal Academy of Sciences in the Netherlands.
\end{acknowledgments}

\end{document}